\documentclass{article}
\pdfoutput=1

\usepackage{arxiv}
\usepackage{amsmath}

\usepackage{algorithm}
\usepackage{algcompatible}
\usepackage{adjustbox}
\usepackage{amsfonts}
\usepackage{amssymb}
\usepackage{amsbsy}
\usepackage[T1]{fontenc}
\usepackage{textcomp}
\usepackage{graphicx}
\usepackage{listings}
\usepackage{multirow}
\usepackage[normalem]{ulem}
\usepackage{color}
\definecolor{ruby}{rgb}{0.6,0,0.3}
\definecolor{gold1}{rgb}{0.8667,0.8510,0.7647}
\definecolor{darkblue}{rgb}{0.0,0.2635,0.4517}
\definecolor{gray1}{rgb}{0.9,0.9,0.9}

\usepackage{soul}
\newcommand{\hlc}[2][gold1]{ {\sethlcolor{#1} \hl{#2}} }
\newcommand{\sourcecode}[2][gray1]{\hspace{-1mm}\hlc[#1]{\small\texttt{#2}}}
\newcommand{\plaincode}[1]{\sourcecode[white]{#1}}
\newcommand{\aquote}[1]{\begin{tabular}{p{2mm}|p{0.5mm}p{140mm}}& & #1\\ \end{tabular}}
\newcommand{\vhref}[2]{\href{#1}{{\color{darkblue}#2}}}

\usepackage{epstopdf}
\usepackage[caption=false]{subfig}

\usepackage{multirow}
\usepackage{xparse}
\ExplSyntaxOn
\NewDocumentCommand{\smallcaps}{m}
 {
  \tl_set:Nn \l_tmpa_tl { #1 }
  \regex_replace_all:nnN
   { ([0-9]+) } 
   { \c{resizedigit}\cB\{ \1 \cE\} } 
   \l_tmpa_tl
  \textsc{ \tl_use:N \l_tmpa_tl }
 }

\ExplSyntaxOff

\usepackage{tikz}

\usepackage[hidelinks]{hyperref}
\usepackage{url}

\graphicspath{{./}}
\newcommand{\ignore}[1]{}
\newcommand{\revisit}[1]{}

\title{An experience with PyCUDA: Refactoring an existing implementation of a ray-surface intersection algorithm}
\shorttitle{An experience with PyCUDA}

\date{May 3, 2023}	

\author{
  \textbf{Raymond~Leung}\vspace{2mm} \\
  Australian Centre for Field Robotics\\
  Faculty of Engineering\\
  The University of Sydney\\ \vspace{-10mm}
}

\renewcommand{\headerleft}{}
\renewcommand{\headeright}{\copyright\,Raymond Leung, 2023}
\renewcommand{\undertitle}{}

\begin{document}
\tikz[remember picture,overlay]
  \node[anchor=north] at (125mm,-30mm)
    {\hspace{35mm}\includegraphics[width=45mm,trim=0mm 0 0 0,clip]{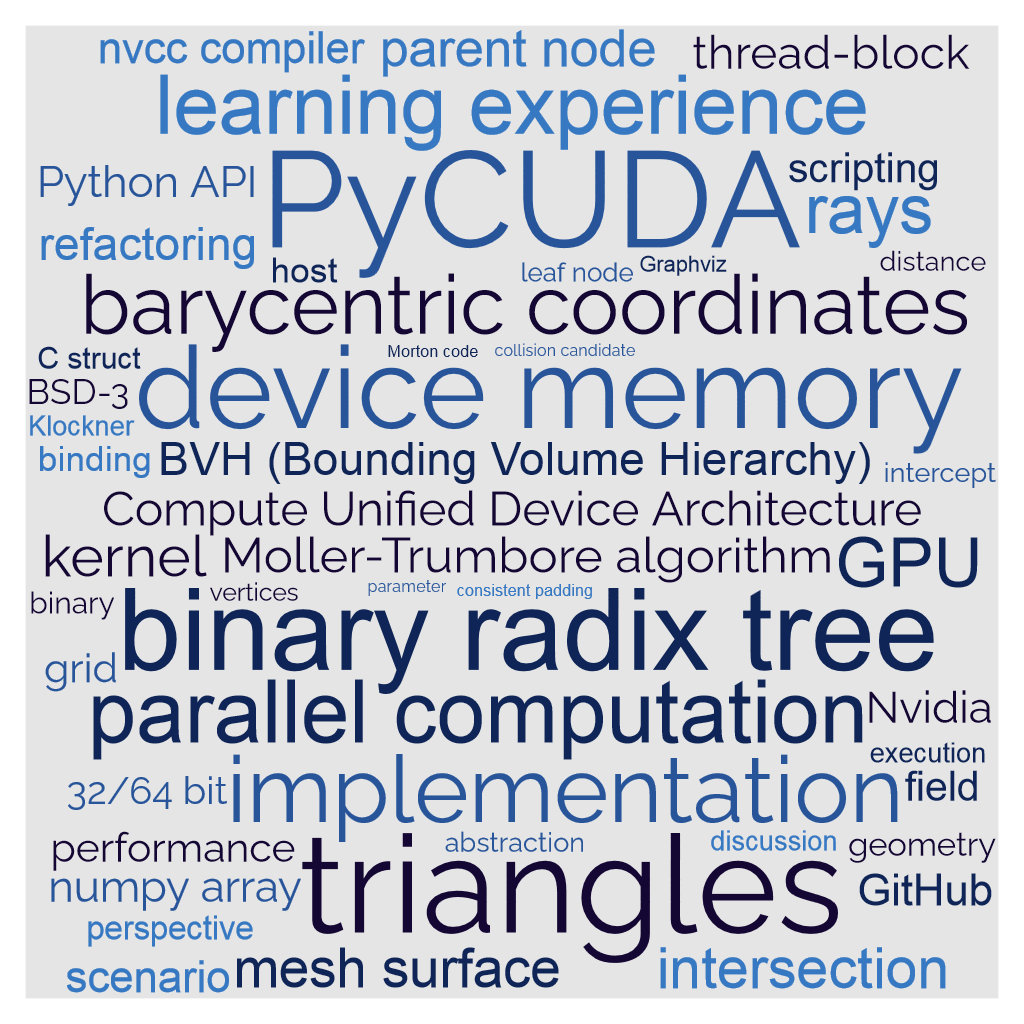}};
\maketitle

\begin{abstract}
This article is a sequel to ``GPU implementation of a ray-surface intersection algorithm in CUDA'' (\vhref{https://arxiv.org/abs/2209.02878}{arXiv:2209.02878}) \cite{leung2022gpursi}. Its main focus is PyCUDA which represents a Python scripting approach to GPU run-time code generation in the Compute Unified Device Architecture (CUDA) framework. It accompanies the open-source code distributed in \vhref{https://github.com/raymondleung8/gpu-ray-surface-intersection-in-cuda/tree/main/pycuda}{GitHub} which provides a PyCUDA implementation of a GPU-based line-segment, surface-triangle intersection test. The objective is to share a PyCUDA learning experience with people who are new to PyCUDA. Using the existing CUDA code and foundation from \cite{leung2022gpursi} as the starting point, we document the key changes made to facilitate a transition to PyCUDA. As the CUDA source for the ray-surface intersection test contains both host and device code and uses multiple kernel functions, these notes offer a substantive example and real-world perspective of what it is like to utilise PyCUDA. It delves into custom data structures such as binary radix tree and highlights some possible pitfalls. The case studies present a debugging strategy which may be used to examine complex C structures (such as a triangle-mesh bounding volume hierarchy) in device memory using standard Python tools without the CUDA-GDB debugger.
\end{abstract}

\section{Background}\label{sec:background}
PyCUDA represents a scripting-based approach to GPU run-time code generation. Its design philosophy and technical elements have previously been covered by its architect and collaborators in \cite{kloeckner-gpu-rtcg09} and \cite{kloeckner-gpu-scripting-pycuda13}. In this paper, PyCUDA discussion is centered around practicalities, and it is told from the perspective of solving a specific geometry problem, where the goal is to determine if a set of rays intersect a triangle mesh surface. For simplicity, we use the term `ray' to refer to a line segment specified by a start and end point; this may be written as $l_i=(\mathbf{r}_{i}^\text{start},\mathbf{r}_{i}^\text{end})$. A typical problem involves a large number of rays, say $N_\text{r}\in[10^6, 10^8]$, which makes parallel computation highly desirable. The mesh surface contains $N_\text{t}$ triangles (typically $N_\text{t}\sim 10^4$), each described by a triplet $\mathbf{t}_j=[t_{j,1},t_{j,2},t_{j,3}]$ which references the three vertices $\mathbf{v}(\mathbf{t}_j)\equiv[\mathbf{v}_{t_{j,1}},\mathbf{v}_{t_{j,2}},\mathbf{v}_{t_{j,3}}]$ from the collection $\{\mathbf{v}_n\}_{1\le n\le N_\text{v}}$. The Moller-Trumbore algorithm \cite{moller1997fast} provides an efficient solution to this problem. However, an exhaustive search [comparing $N_\text{t}$ triangles with $N_\text{r}$ rays] is very slow if acceleration structures are not used even when computation is done on a GPU (see Table 1 in \cite{leung2022gpursi}).

In the existing CUDA implementation, Morton code is used to encode the location of all mesh triangles. These are subsequently reordered to preserve spatial proximity and represented by a bounding volume hierarchy (BVH) using a binary radix tree \cite{karras2012maximizing}. This allows a small subset of triangles [collision candidates] that could possibly intersect with each ray to be quickly identified via tree search. As an overview, the general work flow is shown in {\color{darkblue}Figure~\ref{fig:flowchart}}. Core elements of the implementation are described in Sections 2 and 3 in \cite{leung2022gpursi} and may be cross-referenced in \vhref{https://github.com/raymondleung8/gpu-ray-surface-intersection-in-cuda}{GitHub}.\footnote{Existing CUDA implementation: \url{https://github.com/raymondleung8/gpu-ray-surface-intersection-in-cuda}.}

\begin{figure}[!htb]
\centering
\includegraphics[width=0.8\textwidth]{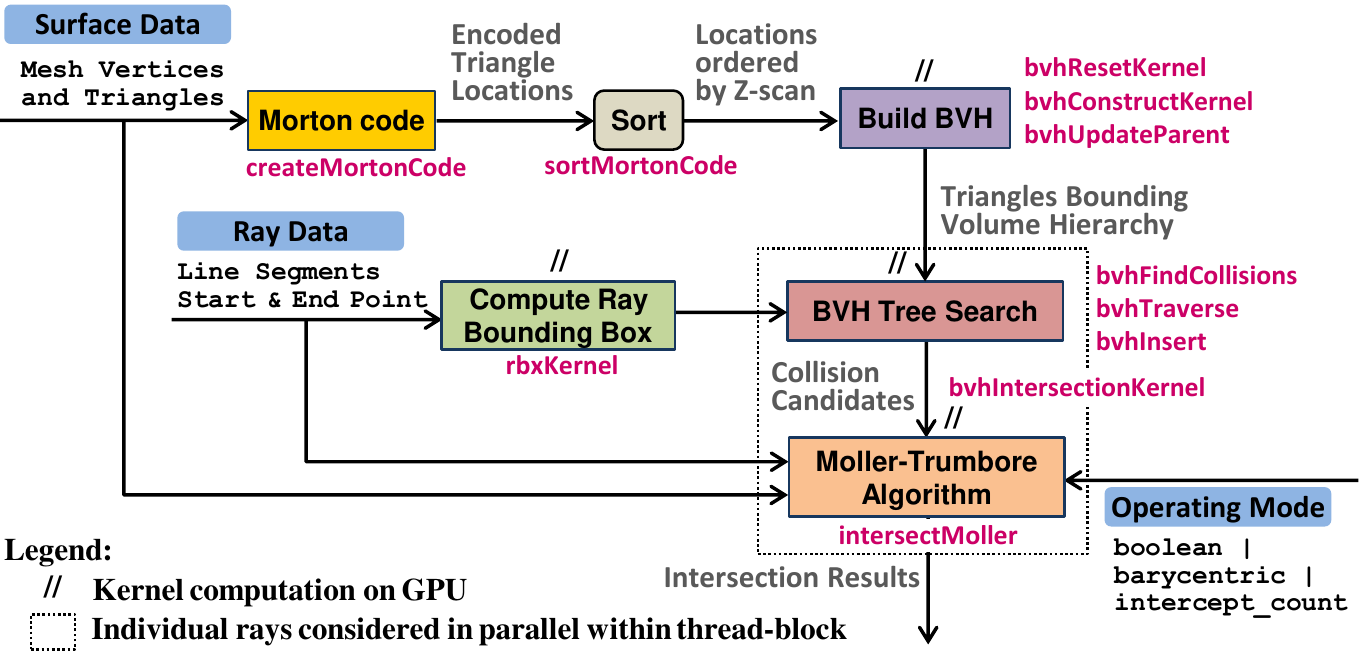}\\ \vspace{5mm}
\includegraphics[width=0.8\textwidth]{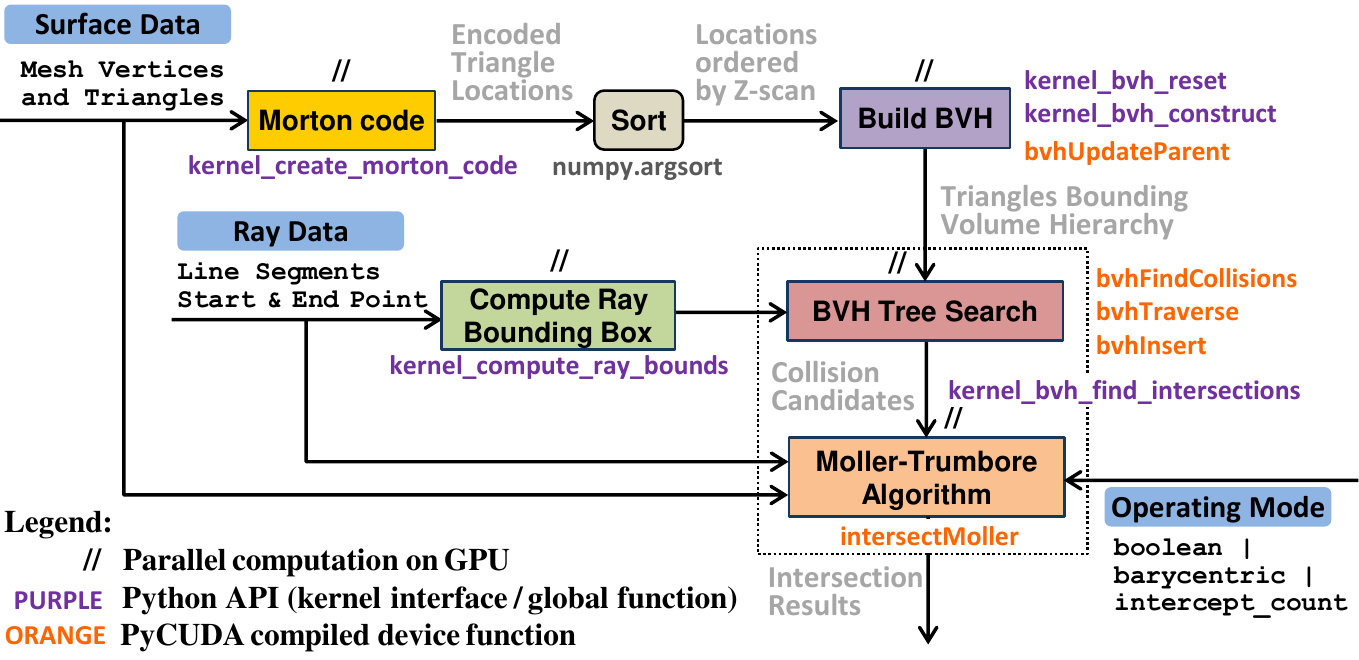}
\caption{Ray-surface intersection detection work flow in the (top) CUDA and (bottom) PyCUDA implementation} \label{fig:flowchart}
\end{figure}

The required inputs and computational components are depicted in the diagram. Parallel processes (kernel codes) are labelled with the // symbol. The program may be configured to operate in different modes to support three use cases.
\begin{itemize}
\item Standard usage ({\small\texttt{mode=boolean}}) --- Return results as a ($N_\text{r},1$) boolean array indicating whether or not each ray intersects with the surface.
\item Extended usage ({\small\texttt{mode=barycentric}}) --- Returns the intersecting rays with the intersecting triangle, the \nobreak{distance} to and location of the nearest intersecting point.
\item Experimental feature ({\small\texttt{mode=intercept\_count}}) --- Returns the number of unique intersections with the surface for each ray.
\end{itemize}

The rest of this article is organised as follows. Section~\ref{sec:prelim} outlines PyCUDA's core features and installation steps. It includes a usage example of the new PyCUDA library. Section~\ref{sec:implementation} focuses on implementation issues and describes the changes that made the existing CUDA code work in PyCUDA. Section~\ref{sec:lessons-learned} presents the lessons learned and describes a debugging approach which may be used to inspect C structures that reside in device memory using Python tools.

\section{Preliminaries}\label{sec:prelim}
\subsection{PyCUDA: What is it and why use it?}\label{sec:pycuda-why}
As mentioned in the introduction, PyCUDA provides a Python integrated approach to GPU code generation at run-time. PyCUDA gives the programmer Pythonic access to Nvidia's parallel computation API \cite{kloeckner-pycuda-doc}. It allows raw CUDA code to be embedded as a string \!\sourcecode{s}\! within a Python script, parsed by \!\sourcecode{pycuda.compilerSourceModule}\!, compiled using nvcc (the CUDA compiler driver) and linked against the CUDA runtime library. The command \!\sourcecode{compilerSourceModule(s)}\! returns a module object \!\plaincode{m}\! which exposes CUDA kernels (\!\plaincode{\_\_global\_\_}\! functions) to the user via \!\sourcecode{m.get\_function(<cuda\_kernel\_name>)} using the \plaincode{cuModuleGetFunction} API.

\aquote{In CUDA terminology, a \plaincode{\_\_global\_\_} function represents kernel code that is executed in an Nvidia GPU. All dependent functions called by the kernel are declared with the \plaincode{\_\_device\_\_} specifier.}

Several features may be noted based on the PyCUDA \vhref{https://documen.tician.de/pycuda/}{documentation}
\begin{itemize}
\item \textit{Resource management} is tied to lifetime of objects (akin to RAII idiom in C++). PyCUDA knows about dependencies, it won't detach from a context before all memory allocated in it is also freed.
\item \textit{Automatic error checking} --- PyCUDA is strict about argument types and Boost python bindings for kernel APIs. CUDA errors are translated into Python exceptions.
\item \textit{Numpy.array like abstraction} using \plaincode{pycuda.gpuarray.GPUArray} for vector operations\,/\,basic arithmetic.
\item \textit{Accessible} --- The kernel APIs of interest are completely at the user's disposal.
\item \textit{Speed} --- PyCUDA's base layer is written in C++, so it has negligible impact on throughput performance.
\end{itemize}

In the writer's opinion, PyCUDA can make device code easier to debug. More about this in Section~\ref{sec:lessons-learned}.

\subsubsection{PyCUDA versus CuPy and Numba CUDA}
PyCUDA is designed for CUDA developers who want to integrate code [already] written in CUDA with Python. For machine learning developers who simply want their NumPy-based code to run on GPUs, \vhref{https://docs.cupy.dev/en/stable/overview.html}{CuPy} offers an alternative. It allows users to benefit from fast GPU computation without learning CUDA syntax \cite{nishino2017cupy}.

For people willing to embrace CUDA who prefer programming using Python constructs, there is \vhref{https://numba.pydata.org/numba-doc/latest/cuda/overview.html}{Numba for CUDA GPUs} which retains CUDA concepts, but translates a restricted subset of Python code into CUDA kernels and device functions following the CUDA execution model. In the case of PyCUDA, these kernels and device functions are written in the CUDA/C language; so being less restrictive and readily deployable in CUDA might be an advantage.

\subsubsection{Working with two versions of the source at the same time}\label{sec:work-with-different-versions-at-the-same-time}
To compare the results produced by two different versions of the source, we can work with both at the same time without needing to switch between git branches, compiling and testing the code one after the other. Instead, the following workflow may be used within a Python environment.\footnote{For simplicity, the compiler cache is not shown in this figure. Refer to caching of GPU binaries (see Fig.\,2) in \cite{kloeckner-gpu-rtcg09}.} This could be useful when we need assurance that incremental changes do not alter the test outcomes during development. The results may be compared using numpy functions and any differences can be easily identified. This scenario is shown in  \plaincode{run\_example3} in \vhref{https://github.com/raymondleung8/gpu-ray-surface-intersection-in-cuda/tree/main/pycuda/demo.py}{\plaincode{pycuda/demo.py}}
\begin{figure}[!htb]
\centering
\includegraphics[width=0.8\textwidth]{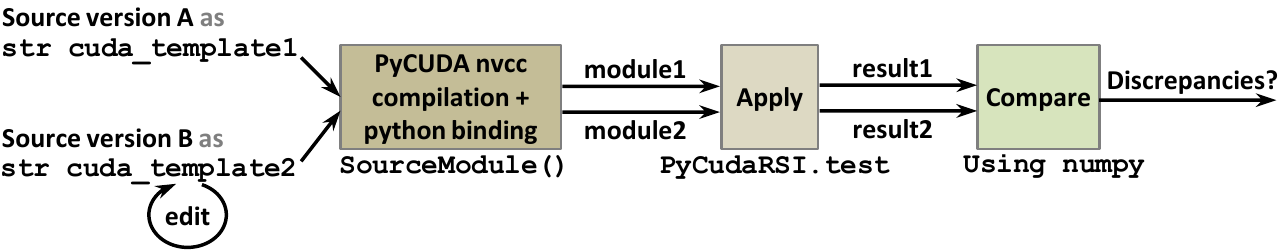}\\
\caption{Comparing results from different source versions within the PyCUDA environment} \label{fig:flowchart}
\end{figure}

In \vhref{https://github.com/raymondleung8/gpu-ray-surface-intersection-in-cuda/tree/main/pycuda/pycuda_ray_surface_intersect.py}{\plaincode{pycuda\_ray\_surface\_intersect.py}}, the \plaincode{try} clause in \plaincode{perform\_bindings\_} and conditional statements associated with \plaincode{self.kernel\_intersect\_*} are used jointly to manage an asymmetric situation where kernels absent from the legacy code (version A of the CUDA module) are introduced in the \plaincode{development} code (version B).

\subsection{PyCUDA installation}\label{sec:pycuda-install}
The prerequisites are listed in the official PyCUDA website. In essence, PyCUDA requires a\vspace{-2mm}
\begin{itemize}
\item \vhref{https://docs.nvidia.com/cuda/cuda-c-programming-guide/index.html\#compute-capabilities}{CUDA-capable Nvidia GPU}
\item Working \href{https://www.python.org/}{Python installation} (in our case, version 3.8)
\item C++ compiler preferrably \vhref{https://gcc.gnu.org/}{GCC} and
\item Access to Nvidia's \vhref{https://developer.nvidia.com/cuda-toolkit}{CUDA toolkit}
\end{itemize}

Several ENVIRONMENT variables should be set before pip install is attempted. For the author, the following commands resulted in a successful installation.

\begin{lstlisting}[backgroundcolor=\color{gold1}, basicstyle={\ttfamily\footnotesize}, escapechar=@, frame=single, linewidth=180mm]
#!/bin/bash

$ export PATH=/usr/local/cuda-11.2/bin${PATH:+:${PATH}}
$ export LD_LIBRARY_PATH=/usr/local/cuda-11.2/lib64${LD_LIBRARY_PATH:+:${LD_LIBRARY_PATH}}
$ export CUDA_INC_DIR=/usr/local/cuda-11.2/include
$ pip3 install pycuda
\end{lstlisting}

The Python packages in my virtual environment are shown in \vhref{https://github.com/raymondleung8/gpu-ray-surface-intersection-in-cuda/tree/main/pycuda/requirements.txt}{pycuda/requirements.txt}.

\subsection{PyCUDA ray-surface intersection library: Usage}\label{sec:pycuda-usage}
The new PyCUDA library (not to be confused with the existing CUDA implementation in \cite{leung2022gpursi}) is used in the following manner. The valid options for \plaincode{mode} are \plaincode{'boolean'}, \plaincode{'barycentric'} and \plaincode{'intercept\_count'}.

\begin{lstlisting}[backgroundcolor=\color{gray1}, basicstyle={\ttfamily\footnotesize}, escapechar=@, frame=single, linewidth=165mm]
from pycuda_ray_surface_intersect import PyCudaRSI

design_params = dict()
cfg = {'mode': mode}

with PyCudaRSI(design_params) as pycu:
    if mode != 'barycentric':
        ray_intersects = pycu.test(vertices, triangles, raysFrom, raysTo, cfg)
    else:
        (intersecting_rays, distances, hit_triangles, hit_points) \
            = pycu.test(vertices, triangles, raysFrom, raysTo, cfg) 
\end{lstlisting}
Refer to \vhref{https://github.com/raymondleung8/gpu-ray-surface-intersection-in-cuda/tree/main/pycuda/demo.py}{pycuda/demo.py} for further comments and diagnostic features.

\subsection{PyCUDA ray-surface intersection: Source Code}
For reference, the PyCUDA ray-surface intersection library is available at
\vspace{-2mm}\begin{itemize}
\item[] {\color{blue}\url{https://github.com/raymondleung8/gpu-ray-surface-intersection-in-cuda/tree/main/pycuda/}}
\end{itemize}
\vspace{-2mm}It is distributed open-source under the BSD 3-clause license.

\section{PyCUDA implementation}\label{sec:implementation}
The core functionalities are implemented in \sourcecode{pycuda\_ray\_surface\_intersect.py}. Within this script, the C++/CUDA source code is retrieved from \sourcecode{pycuda\_source.py} using \sourcecode{get\_cuda\_template()}. It returns a module string which contains nearly the same code as found in \plaincode{"rsi\_geometry.h"}, \plaincode{"morton3D.h"} and \plaincode{"bvh\_structure.h"}. This separation is not technically necessary, but it makes the the class {\color{ruby}\plaincode{PyCudaRSI}} in \plaincode{pycuda\_ray\_surface\_intersect.py} more readable. \plaincode{PyCudaRSI} (more or less) replicates the code found in \plaincode{gpu\_ray\_surface\_intersect.cu} (the existing CUDA implementation from \cite{leung2022gpursi}) and provides user with high-level interfaces that support the three primary use cases.

\subsection{Refactoring existing CUDA code for PyCUDA}\label{sec:refactor-cuda-for-pycuda}
By default, PyCUDA sets \plaincode{no\_extern\_c = False}. This has the effect of wrapping the entire source code specified in the \plaincode{cuda\_template} with \plaincode{extern "C" \{\}} which prevents C++ \vhref{https://en.wikipedia.org/wiki/Name_mangling}{name mangling} and allows the APIs of interest to be identified. This means, if the \sourcecode{\#include <vector>} statement from \plaincode{"bvh\_structure.h"} is retained [note: STL is only permitted in host code, it is never used in device code], it would trigger a linkage error like

\aquote{include/c++/6.2.1/bits/stl\_iterator\_base\_types.h(116):}
\aquote{Error: this declaration may not have extern "C" linkage.}

PyCUDA also does not support overloaded functions. Although it allows templated \plaincode{\_\_device\_\_} functions to be used, templated kernel functions (like \!\plaincode{template <typename M>
\_\_global\_\_ void bvhConstruct}\!) appear to be unsupported.\revisit{https://github.com/cupy/cupy/issues/3185#issuecomment-649064011}\footnote{In contrast, CuPy supports \vhref{https://docs.cupy.dev/en/stable/user_guide/kernel.html\#type-generic-kernels}{type-generic kernels} and \vhref{https://docs.cupy.dev/en/stable/user_guide/kernel.html\#raw-modules}{C++ templated kernels} where the un-mangled kernel name and template parameters are unambigously specified using \plaincode{name\_expressions}\!. Name mangling and demangling are handled under the hood.} These resulted in some changes that differ from the \href{https://github.com/raymondleung8/gpu-ray-surface-intersection-in-cuda/}{original CUDA code}.

\subsection{Specific changes}\label{sec:specific-changes}
\begin{itemize}
\item From ``bvh\_structure.h'', the host function \plaincode{void inline createMortonCode(const vector<T> \&vertices, const vector<int> \&triangles,...)} is converted into a kernel \sourcecode{\_\_global\_\_ kernelMortonCode(const float* vertices, const int* triangles,...)}.
  \begin{itemize}
  \item This becomes accessible in Python with the binding \sourcecode{kernel\_create\_morton\_code = m.get\_function("kernelMortonCode")} which allows the location of mesh triangles to be transformed into 64-bit Morton codes.
  \end{itemize}
\item Sorting of the Morton codes, which essentially organises the triangles into local clusters to preserve spatial proximity, is no longer performed using a C++ host function. Instead, it is done in python, and the result is subsequently copied to device memory \sourcecode{d\_sortedTriangleIDs}.
\item From ``morton3D.h", the \sourcecode{template<typename morton, typename coord>} parameters are removed. Instead, the types \plaincode{MORTON, COORD} in the \sourcecode{cuda\_template} are substituted with concrete types using a dictionary (e.g.\!\plaincode{\{"MORTON": "uint64\_t", "COORD": "unsigned int"\}}\!) before being parsed by \sourcecode{\nobreak{pycuda}.compilerSourceModule(...)}.
\item Finally, the existing CUDA implementation makes extensive use of custom data structures (such as axes-aligned bounding box \sourcecode{AABB}, \sourcecode{BVHNode}, \sourcecode{CollisionList} and \sourcecode{InterceptDistances}). \\For native types like \plaincode{float*}, device memory allocation can be done simply using \sourcecode{d\_vertices = cuda.mem\_alloc(h\_vertices.nbytes)} for instance, where \plaincode{h\_vertices} is a numpy.array of type \plaincode{numpy.float32} and shape \plaincode{(n\_vertices,3)}. For C structures, a helper method \sourcecode{struct\_size(cuda\_szQuery, d\_answer)} is defined to convey the bytesize of the relevant structure to Python. This must be known in order to allocate global memory on the GPU device using PyCUDA.
  \begin{itemize}
  \item Here, \sourcecode{cuda\_szQuery} refers to the Python interface to a getSize function such as \sourcecode{\_\_global\_\_ void bytesInBVHNode(int \&x)\{ x = sizeof(BVHNode); \}}.
  \item The bytesize should not be hardcoded because memory padding may be introduced at compile time, and the extent may vary depending on the device architecture. Using the \plaincode{\_\_align\_\_} compiler specifier (or some other variant) is encouraged to ensure CUDA memory alignment is consistent.
  \end{itemize}
\end{itemize}

Despite these differences, the new code in \plaincode{pycuda\_ray\_surface\_intersect.py} is remarkably similar to the existing CUDA implementation.

\subsection{Performance}\label{sec:performance}
Speed and correctness of the program are always front of mind. In this section, we first look at speed and identify areas where it can be improved. The following table compares the execution time for the original CUDA implementation \cite{leung2022gpursi} with PyCUDA. For PyCUDA, the elapsed time (see \!\plaincode{t\_start} and \plaincode{t\_end} in \plaincode{pycuda\_ray\_surface\_intersect.py}\!) also includes I/O overhead (e.g.\!\plaincode{memcpy\_dtoh}\!operations orchestrated by PyCUDA) and numpy processing when run in \plaincode{barycentric} mode. In contrast, the original CUDA program writes results to binary files; derived quantities (e.g. intersecting points) are not computed. Overall, run times are comparable in \!\plaincode{boolean}\! and \!\plaincode{intercept\_count}\! mode.

\begin{lstlisting}[backgroundcolor=\color{white}, basicstyle={\ttfamily\footnotesize}, escapechar=@, frame=single, linewidth=165mm]
 Test data generated by "diagnostic_input.py" contains
 10M rays and a surface with 14874 vertices, 29260 triangles.

 | Operating mode | Original CUDA code | PyCUDA implementation |
 |----------------|--------------------|-----------------------|
 | boolean        |      300.166 ms    |       355.094 ms      |
 | barycentric    |      308.625 ms    |       909.376 ms      |
 | intercept_count|      334.192 ms    |       383.117 ms      |
\end{lstlisting}

The next table provides a more detailed breakdown in \plaincode{barycentric} mode.
\begin{lstlisting}[backgroundcolor=\color{white}, basicstyle={\ttfamily\footnotesize}, escapechar=@, frame=single, linewidth=165mm]
 | Component   Description                       | Duration    |
 |-----------------------------------------------|-------------|
 | 1 | Core processing*                          | 138.513 ms  |
 | 2 | memcpy_dtoh (intersectTriangles, baryT)   | 231.357 ms  |
 | 3 | Calculating intersecting points/distances |  @{\color{ruby}539.506 ms}@  |
 |-------------------------------------------------------------|
 | 3a| Identify intersecting rays (numpy)        |  34.562 ms  |
 | 3b| Calculate intersecting distances (host)   | 240.836 ms  |
 | 3c| Retain intersecting triangles    (host)   |   9.246 ms  |
 | 3d| Calculate intersecting points    (host)   | 254.862 ms  |

 * From@\sourcecode{t\_start}@to completion of@\sourcecode{kernel\_bvh\_find\_intersections2}@.
   This includes: configure_, allocate_memory_, transfer_data_,
   get_min_max_extent_of_surface, kernel_compute_ray_bounds,
   kernel_create_morton_code, kernel_bvh_reset, kernel_bvh_construct.
\end{lstlisting}

Based on Occam's principle, the host code for components 3b and 3d were converted to kernel code to increase efficiency. These kernel functions (\!\plaincode{kernelIntersectDistances}\! and \!\plaincode{kernelIntersectPoints}\!) were added to \plaincode{\nobreak{pycuda}\_source.py}. With minimal effort, significant improvement can be seen from the run time figures below.
\begin{lstlisting}[backgroundcolor=\color{white}, basicstyle={\ttfamily\footnotesize}, escapechar=@, frame=single, linewidth=165mm]
 | Component   Description                       | Duration    |
 |-----------------------------------------------|-------------|
 | 3b| Calculate intersecting distances (gpu)    |  18.039 ms  |
 | 3d| Calculate intersecting points    (gpu)    |  24.629 ms  |
 |-------------------------------------------------------------|
 | 3 | Calculating intersecting points/distances |  86.476 ms  | (84% reduction)
 | * | Total duration for barycentric            |  @{\color{blue}456.346 ms}@   | (50% reduction)
\end{lstlisting}
Next, we will take a deep-dive and discuss debugging and verification efforts.

\section{Lessons learned}\label{sec:lessons-learned}
A significant advantage is that PyCUDA can make device code easier to debug. This in part is due to the ease with which device memory can be copied to the host for inspection as numpy.arrays. However, there are also pitfalls that developers need to be mindful of. In the ensuing discussion, we examine two situations: one concerns abnormal test results, another resulted in the program crashing due to grid partitioning (carelessness in the binary radix tree construction). Both issues had their origins in memory use, which may be diagnosed by examining contents in the triangle-mesh \vhref{https://en.wikipedia.org/wiki/Bounding_volume_hierarchy}{bounding volume hierarchy}.

\subsection{Case Study 1: Debugging C struct from device memory. Meaning of `int' in CUDA and numpy}\label{sec:case-study1}
\vspace{-5mm}
\begin{figure}[h]
\centering
\includegraphics[width=65mm]{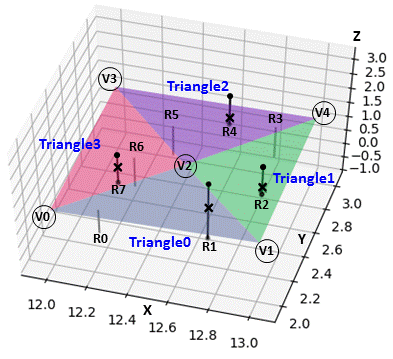}\vspace{-2mm}
\caption{A simple test configuration with 8 rays and 4 triangles} \label{fig:simple-test}
\end{figure}

This picture depicts a test scenario where the mesh surface consists of 4 triangles and there are 8 rays (line segments). In particular, R\textsubscript{1}, R\textsubscript{2}, R\textsubscript{4} and R\textsubscript{7} intersect triangle T\textsubscript{0}, T\textsubscript{1}, T\textsubscript{2} and T\textsubscript{3} at [12.7,2.2,1.14], [12.9,2.4,1.21], [12.6,2.9,1.23] and [12.2,2.4,1.08], respectively. However, due to a bug, the initial PyCUDA implementation reported only R\textsubscript{1} and R\textsubscript{7} as the intersecting rays. Through a process of elimination, the cause was narrowed down to BVH tree construction and interpretation of the triangle IDs after the corresponding Morton codes are sorted.

To diagnose the problem, the contents of the triangle bounding volume hierarchy (BVH) were examined. The aim is to verify the integrity of the BVH and find the root cause. This involves two steps as listed below.

\subsubsection{Bounding Volume Hierarchy (BVH) Debugging Strategy}\label{sec:bvh-debug-strategy}
\begin{enumerate}
\item Transfer the relevant content from device memory to host memory using standard PyCUDA I/O interfaces.
\item Decode the fields within each BVH node based on the type definitions given in \sourcecode{struct BVHNode}.
\end{enumerate}
In our current circumstances, the goal is to flush out all information and examine some properties (e.g. connectivity) within a graph. So, this strategy offers a suitable alternative to invoking \plaincode{cuda-gdb} as many of its rich features are not needed. It is certainly not intended as a replacement of \plaincode{cuda-gdb} in the general sense.

\begin{figure}[h]
\centering
\includegraphics[width=120mm]{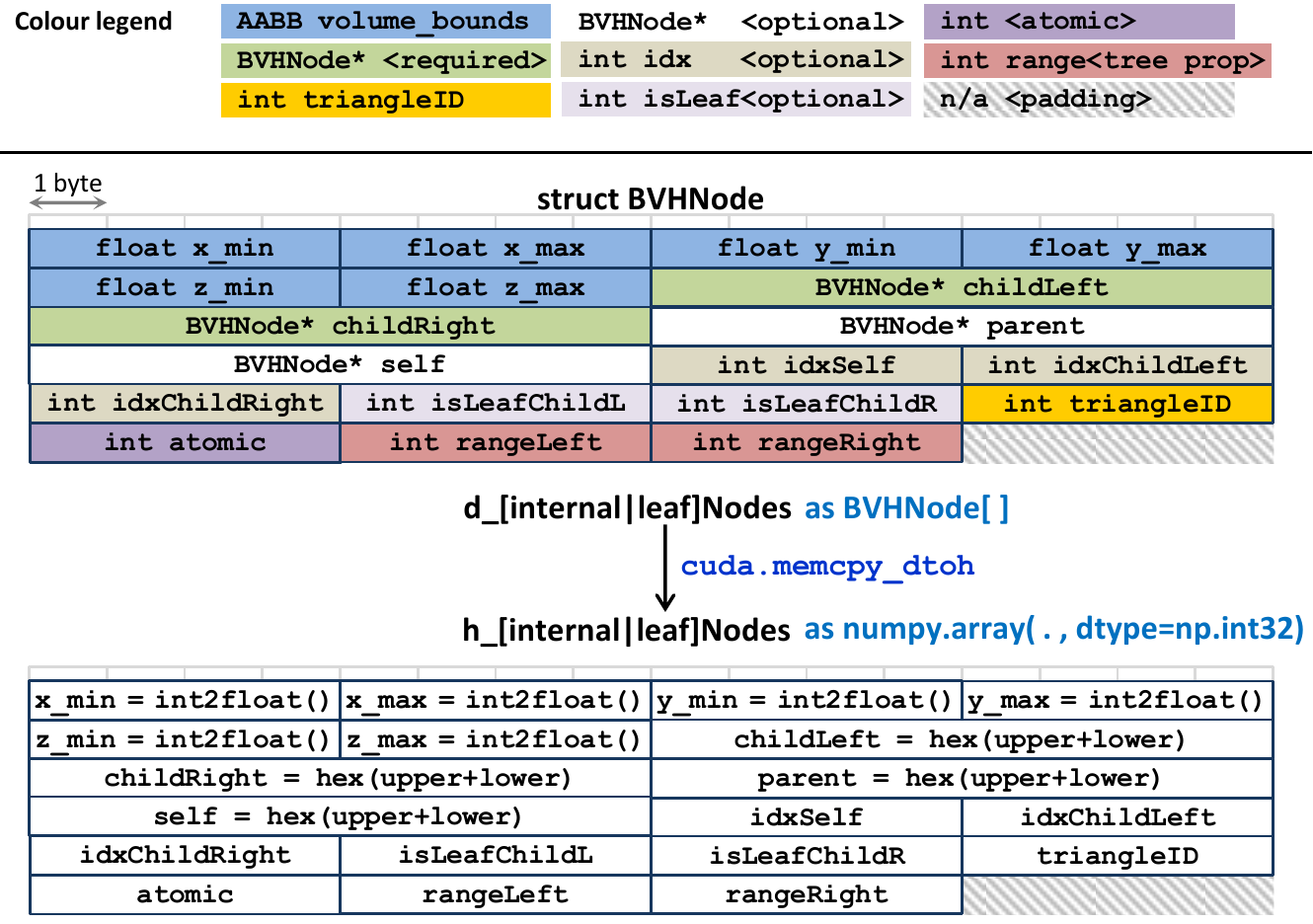}
\caption{A bounding volume hierarchy consists of $N_\text{t}\!$ internal nodes (the last one only holds a pointer to the root node) and $N_\text{t}$ leaf nodes. Each memory array is copied from GPU device to host (using int32 representation). The contents within individual BVHNodes are subsequently decoded according to the data type of each field.} \label{fig:bvh-structure}\end{figure}

Specifically, two numpy arrays are created to hold the content of the internal and leaf nodes. The \plaincode{cuda.memset\_dtoh} operation converts the bitstream into \plaincode{int32} values.

\begin{lstlisting}[backgroundcolor=\color{gray1}, basicstyle={\ttfamily\footnotesize}, escapechar=@, frame=single, linewidth=170mm]
h_leafNodes = np.zeros(n_triangles * sz_BVHNode, dtype=np.int32)
h_internalNodes = np.zeros(n_triangles * sz_BVHNode, dtype=np.int32)
cuda.memcpy_dtoh(h_leafNodes, d_leafNodes)
cuda.memcpy_dtoh(h_internalNodes, d_internalNodes)
\end{lstlisting}

To manipulate the word-stream (where `word' = 4 contiguous bytes), the memory footprint of each \plaincode{BVHNode} is determined using the helpful \sourcecode{struct\_size} macro [defined in the \plaincode{pycuda\_ray\_surface\_intersect.py} module] and divided by the number of bytes in \plaincode{numpy.int32} to calculate the number of \plaincode{int32} elements per \plaincode{BVHNode}.

\begin{lstlisting}[backgroundcolor=\color{gray1}, basicstyle={\ttfamily\footnotesize}, escapechar=@, frame=single, linewidth=165mm]
sz_BVHNode = int(struct_size(bytes_in_BVHNode, d_szQuery)
               / np.ones(1, dtype=np.int32).nbytes)
\end{lstlisting}

Content within each \plaincode{BVHNode} (a chunk of \sourcecode{h\_*Nodes}\hspace{-1mm}) is processed one node at a time using the \sourcecode{display\_node\_contents} method in \sourcecode{pycuda/diagnostic\_utils.py} which interprets the field values according to the definition of \plaincode{struct BVH\,Node}. For debugging purpose, \sourcecode{params['USE\_EXTRA\_BVH\_FIELDS']} is set to \plaincode{True} to (i) indicate whether the Left/Right descendants (\plaincode{childL} and \plaincode{childR} nodes) are "internal" or "leaf" (terminal) nodes; and (ii) include the memory address of the parent and current node itself, to make the binary radix tree structure easier to trace.

\begin{lstlisting}[backgroundcolor=\color{gray1}, basicstyle={\ttfamily\footnotesize}, escapechar=@, frame=single, linewidth=165mm]
for t in range(n_triangles):
    words = h_internalNodes[t*sz_BVHNode:(t+1)*sz_BVHNode]
    display_node_contents(words, t, params['USE_EXTRA_BVH_FIELDS'])

for t in range(n_triangles):
    words = h_leafNodes[t*sz_BVHNode:(t+1)*sz_BVHNode]
    display_node_contents(words, t, params['USE_EXTRA_BVH_FIELDS'])
\end{lstlisting}

Distilling the critical information, the code above reveals that the first triangle T\textsubscript{0} was repeated in the leaf nodes. Furthermore, aside from T\textsubscript{3}, the triangle IDs for T\textsubscript{1} and T\textsubscript{2} were also missing.

\begin{lstlisting}[backgroundcolor=\color{white}, basicstyle={\ttfamily\footnotesize}, escapechar=@, frame=single, linewidth=165mm]
Leaf nodes
[0] x:[12.0,13.0], y:[2.0,2.5], z:[1.0,1.2], triangleID: @{\color{ruby}0}@, rangeL: 0, rangeR: 0
[1] x:[12.0,13.0], y:[2.0,2.5], z:[1.0,1.2], triangleID: @{\color{ruby}0}@, rangeL: 1, rangeR: 1
[2] x:[12.0,12.5], y:[2.0,3.0], z:[1.0,1.2], triangleID: @{\color{ruby}3}@, rangeL: 2, rangeR: 2
[3] x:[12.0,13.0], y:[2.0,2.5], z:[1.0,1.2], triangleID: @{\color{ruby}0}@, rangeL: 3, rangeR: 3
\end{lstlisting}

\subsubsection{The culprit}\label{sec:culprit1}
In \sourcecode{kernelBVHReset}, the argument \plaincode{sortedObjectIDs} is of type \sourcecode{int*}. In CUDA, this is treated as an \sourcecode{int32[]} array.

\begin{lstlisting}[backgroundcolor=\color{gray1}, basicstyle={\ttfamily\footnotesize}, escapechar=@, frame=single, linewidth=165mm]
__global__ void kernelBVHReset(const float* __restrict__ vertices,
                               const int* __restrict__ triangles,
                               BVHNode* __restrict__ internalNodes,
                               BVHNode* __restrict__ leafNodes,
                               int* __restrict__ sortedObjectIDs, int nNodes)
{
    unsigned int i = blockIdx.x * blockDim.x + threadIdx.x;
    if (i >= nNodes)
        return;
    //set triangle attributes in leaf
    leafNodes[i].triangleID = t = sortedObjectIDs[i];
    :
}
\end{lstlisting}
This is incompatible with
\begin{lstlisting}[backgroundcolor=\color{gray1}, basicstyle={\ttfamily\footnotesize}, escapechar=@, frame=single, linewidth=180mm]
d_sortedTriangleIDs = cuda.mem_alloc(n_triangles * np.ones(1, dtype=np.int64).nbytes) #(**)
\end{lstlisting}
which is passed as the \plaincode{sortedObjectIDs} argument. Its dtype follows the convention of
\begin{lstlisting}[backgroundcolor=\color{gray1}, basicstyle={\ttfamily\footnotesize}, escapechar=@, frame=single, linewidth=165mm]
h_sortedTriangleIDs = np.argsort(h_morton) #(**)
\end{lstlisting}
which under Python3, the returned array type \plaincode{int} is defaulted to \sourcecode{int64} on 64-bit machines.

As a result, using \plaincode{p[i]} as a short-hand for \plaincode{d\_sortedObjectIDs[i]}, the figure below shows that the CUDA int32 pointer is incremented at half the required rate (advancing by 32-bit each time). For the third ``read'', \plaincode{p[2]} is fetching the least-significant word of the second 64-bit integer which happens to reference triangle T\textsubscript{3}.
\begin{center}
\includegraphics[width=150mm,trim={0 28mm 0 0},clip]{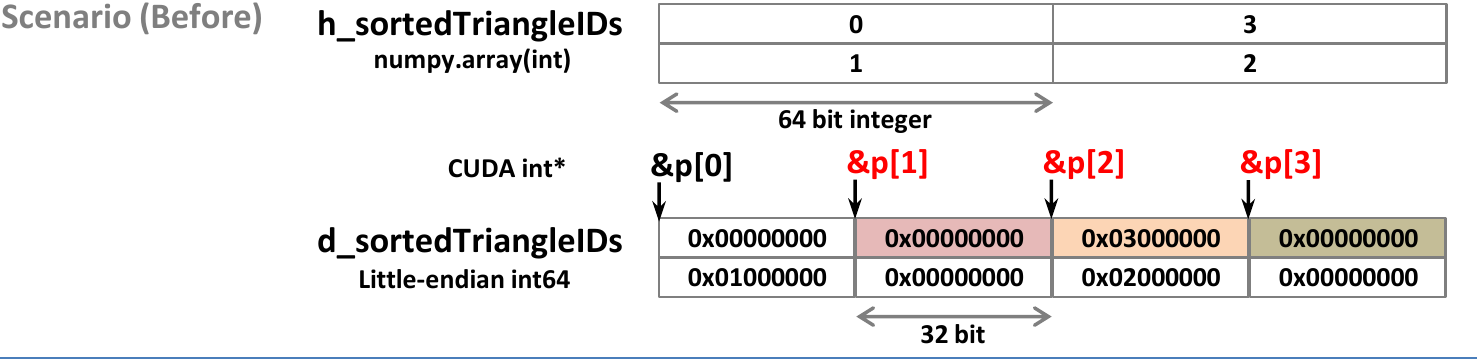}
\end{center}
\begin{figure}[h]
\centering
\includegraphics[width=150mm,trim={0 0 0 38mm},clip]{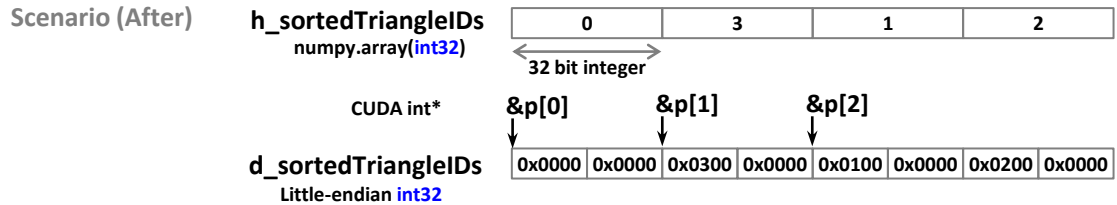}
\caption{The incongruous nature of memory access when CUDA int* reads off 32-bit integers from a 64-bit array. This situation arised because memory for the device array was allocated based on the size of the host array returned by numpy.argsort which interprets int as int64. Meanwhile in CUDA, int is treated as int32 by the compiler
.} \label{fig:int-pointer}
\end{figure}

\newpage
\subsubsection{Resolution}\label{sec:resolution}
To fix this, the stride (for a 32-bit integer pointer in CUDA) needs to be consistent with the width of an integer element in device memory. This may be enforced by replacing the lines (**) above with
\begin{lstlisting}[backgroundcolor=\color{gray1}, basicstyle={\ttfamily\footnotesize}, escapechar=@, frame=single, linewidth=165mm]
h_sortedTriangleIDs = np.argsort(h_morton).astype(np.int32)
d_sortedTriangleIDs = cuda.mem_alloc(n_triangles * np.ones(1,dtype=np.int32).nbytes)
\end{lstlisting}

The final PyCUDA implementation contains these changes and produces debugging output where each of the LEAF nodes corresponds to a different triangle.
\begin{lstlisting}[backgroundcolor=\color{white}, basicstyle={\ttfamily\footnotesize}, escapechar=@, frame=single, linewidth=165mm]
BVH tree structure
---------------------------
Internal nodes
[0] x:[12.0,13.0], y:[2.0,3.0], z:[1.0,1.2]  <<< Bounding Volume
self: 0x66e1200, parent: 0x66e1260
indices: 0(self), 0(L-leaf), 1(R-leaf)
childL: 0x66e1000, childR: 0x66e1060
atomic: 2, rangeL: 0, rangeR: 1

[1] x:[12.0,13.0], y:[2.0,3.0], z:[1.0,1.3]  ------ ROOT NODE
self: 0x66e1260, parent:
indices: 1(self), 0(L-internal), 2(R-internal)
childL: 0x66e1200, childR: 0x66e12c0
atomic: 2, rangeL: 0, rangeR: 3

[2] x:[12.0,13.0], y:[2.0,3.0], z:[1.1,1.3]
self: 0x66e12c0, parent: 0x66e1260
indices: 2(self), 2(L-leaf), 3(R-leaf)
childL: 0x66e10c0, childR: 0x66e1120
atomic: 2, rangeL: 2, rangeR: 3

[3] x:[0.0,0.0], y:[0.0,0.0], z:[0.0,0.0]
self: 0x0, parent: 0x0
indices: 3(self), 1(L-internal), 0(R-internal)
childL: 0x66e1260(root node), childR: 0x0
atomic: 0, rangeL: 0, rangeR: -1

---------------------------
Leaf nodes
[0] x:[12.0,13.0], y:[2.0,2.5], z:[1.0,1.2]
self: 0x66e1000, parent: 0x66e1200
triangleID: @{\color{blue}0}@

[1] x:[12.0,12.5], y:[2.0,3.0], z:[1.0,1.2]
self: 0x66e1060, parent: 0x66e1200
triangleID: @{\color{blue}3}@

[2] x:[12.5,13.0], y:[2.0,3.0], z:[1.1,1.3]
self: 0x66e10c0, parent: 0x66e12c0
triangleID: @{\color{blue}1}@

[3] x:[12.0,13.0], y:[2.5,3.0], z:[1.1,1.3]
self: 0x66e1120, parent: 0x66e12c0
triangleID: @{\color{blue}2}@
\end{lstlisting}

Furthermore, the program correctly reports R\textsubscript{1}, R\textsubscript{2}, R\textsubscript{4} and R\textsubscript{7} as the intersecting rays under `boolean' mode.
\begin{lstlisting}[backgroundcolor=\color{white}, basicstyle={\ttfamily\footnotesize}, escapechar=@, frame=single, linewidth=165mm]
0: 0      1: 1      2: 1      3: 0      4: 1      5: 0      6: 0      7: 1
\end{lstlisting}

\subsubsection{Test code and Graphviz}\label{sec:testcode-graphviz}
These results may be replicated by running {\color{ruby}\sourcecode{pycuda/demo.py}}. Setting \sourcecode{cfg['bvh\_visualisation'] = ['graph']} in \sourcecode{run\_example1}, it will also generate a graph for the mesh-triangle binary radix tree using graphviz.
\begin{itemize}
\item The relevant code \sourcecode{bvh\_graphviz} is found in \sourcecode{pycuda/diagnostic\_graphics.py}.
\end{itemize}
\begin{figure}[h]
\centering
\vspace{-3mm}\includegraphics[width=50mm,trim={0 5mm 0 3mm},clip]{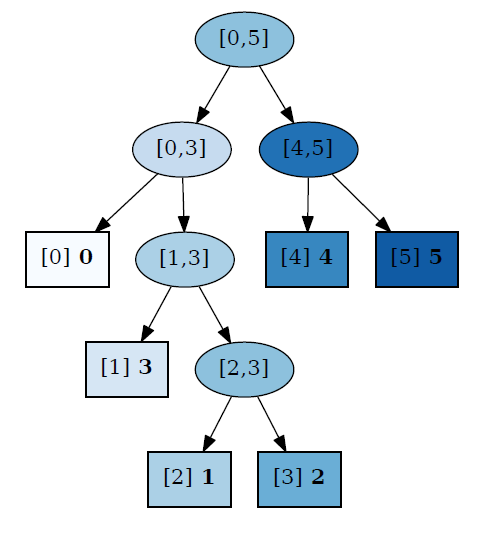}\vspace{-3mm}
\caption{A graph for the binary radix tree generated using Graphviz} \label{fig:bvh-graph1.png}
\end{figure}

The graph above corresponds to the test surface used in \plaincode{intercept\_count} mode, where two patches have been added above triangles 1 and 2 in the simple test surface to form a ``canopy". For the root node and internal nodes, \plaincode{[a,b]} represents the node index range. For terminal nodes (squares), the \plaincode{c} in \plaincode{[c] d} represents the leaf node index and \plaincode{d} represents the triangle ID (taking into account reordering using the Morton codes).
\begin{itemize}
\item This setting also writes a graph description in the DOT language to a file (\plaincode{bvh\_structure.gv}).
\end{itemize}

\subsection{Case Study 2: Not paying enough attention to grid size.\,\linebreak Illegal memory access when BVH structure is not properly constructed}\label{sec:case-study2}
A large surface with $\sim$30,000 triangles and 10 million rays were created using \sourcecode{synthesize\_data} from \plaincode{pycuda/diagnostic\_input.py}. Using similar construction, the \sourcecode{PyCudaRSI.test} API was called in \plaincode{boolean} mode. The following error messages were produced.
\begin{lstlisting}[backgroundcolor=\color{gold1}, basicstyle={\ttfamily\footnotesize}, escapechar=@, frame=single, linewidth=165mm]
Traceback (most recent call last):
  File "/opt/python3.8/lib/python3.8/runpy.py", line 193, in _run_module_as_main
    return _run_code(code, main_globals, None,
  File "/opt/python3.8/lib/python3.8/runpy.py", line 86, in _run_code
    exec(code, run_globals)
  File "<git_repo>/pycuda/demo.py", line 182, in <module>
    run_example2(mode)
  File "<git_repo>/pycuda/demo.py", line 140, in run_example2
    ray_intersects = pycu.test(vertices, triangles, raysFrom, raysTo, cfg)
  File "<git_repo>/pycuda/pycuda_ray_surface_intersect.py", line 313, in test
    cuda.memcpy_dtoh(self.h_crossingDetected, self.d_crossingDetected)
pycuda._driver.LogicError: cuMemcpyDtoH failed:
                           an illegal memory access was encountered
PyCUDA WARNING: a clean-up operation failed (dead context maybe?)
cuModuleUnload failed: an illegal memory access was encountered
\end{lstlisting}

\subsubsection{Investigation}\label{sec:investigation}
To investigate this issue, we examined the tree content once again following the BVH debugging strategy. A print-out is obtained by setting \sourcecode{cfg['examine\_bvh'] = True} temporarily.

\begin{figure}[h]
\centering
\vspace{-3mm}\includegraphics[width=140mm]{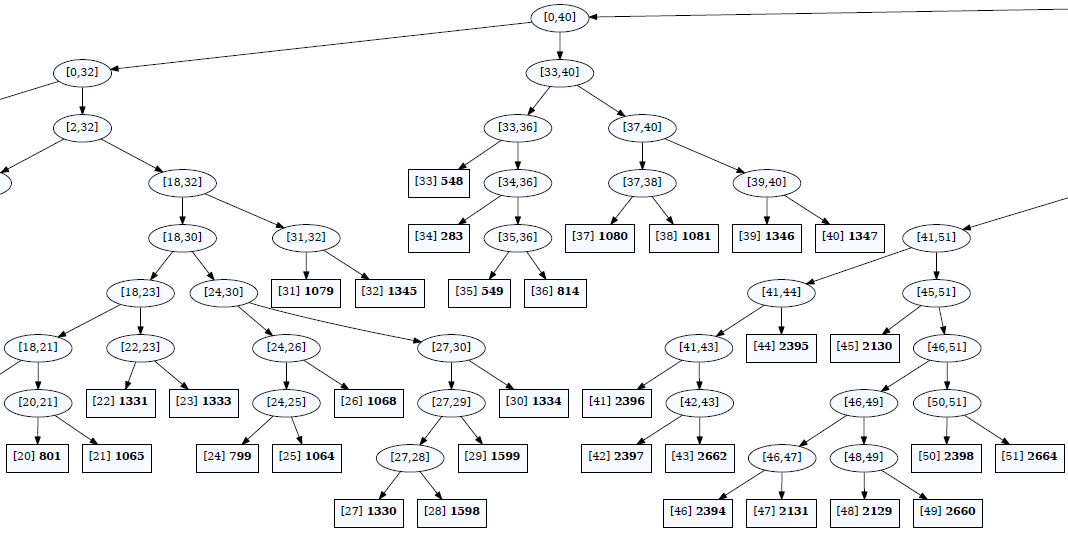}\vspace{-3mm}
\caption{A truncated portion of the graph for the triangle-mesh binary radix tree used in case 2} \label{fig:bvh-graph2}
\end{figure}

In this case, the BVH is much more complex. A truncated portion of the graph is shown in Fig.~\ref{fig:bvh-graph2}.

\begin{figure}[!h]
\centering
\includegraphics[width=110mm,trim={0 0 0 0mm},clip]{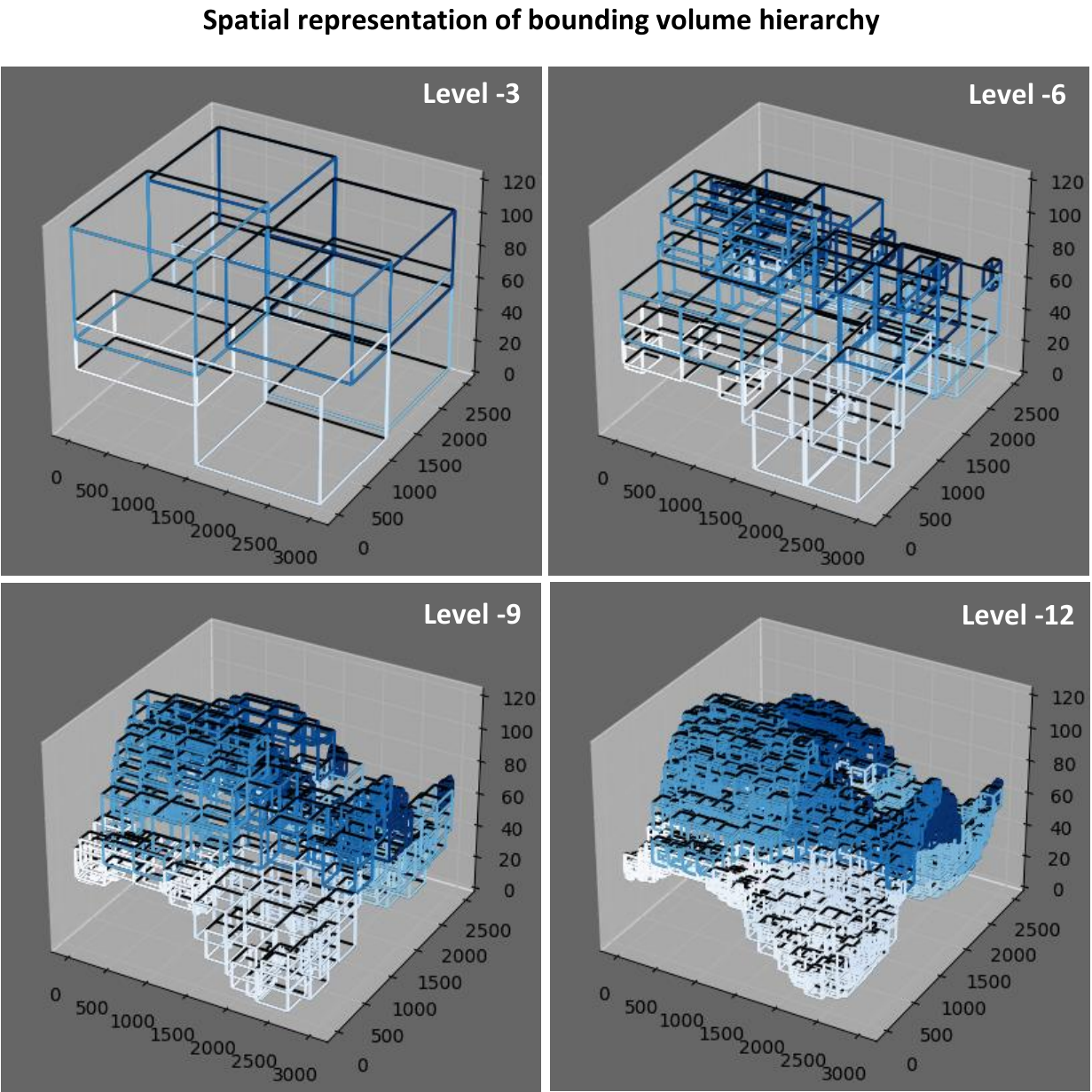}\vspace{-3mm}
\caption{A spatial representation of the bounding volume hierarchy in case 2} \label{fig:bvh-spatial}
\end{figure}

\vspace{-2mm}The next picture (Fig.~\ref{fig:bvh-spatial}) shows what the bounding volume hierarchy looks like as we descend from the top of the tree. The root node is considered level 0, its immediate descendants constitute level -1 and so forth.

\subsubsection{Observations}\label{sec:finding}
The critical findings are included below.
\begin{lstlisting}[backgroundcolor=\color{white}, basicstyle={\ttfamily\footnotesize}, escapechar=@, frame=single, linewidth=165mm]
BVH tree structure
---------------------------
Internal nodes
   :
[16382] x:[992.584,1018.47], y:[363.936,387.745], z:[61.0106,62.1155]
self: 0xe2361ff40, parent: 0xe2361ffa0
indices: 16382(self), 16382(L-leaf), 16383(R-leaf)
childL: 0xe2335ff40, childR: 0xe2335ffa0
atomic: 2, rangeL: 16382, rangeR: 16383

[16383] x:[0,0], y:[0,0], z:[0,0]
self: 0x0, parent: 0x0
indices: 16383(self), 16382(L-internal), 0(R-internal)
childL: 0xe2361ff40, childR: 0x0
atomic: 1, rangeL: 16382, rangeR: -1

[16384] x:[0,0], y:[0,0], z:[0,0]
self: 0x0, parent: 0x0
indices: 16384(self), 0(L-internal), 0(R-internal)
childL: 0x0, childR: 0x0
atomic: 0
   :
[29259] x:[0,0], y:[0,0], z:[0,0]
self: 0x0, parent: 0x0
indices: 29259(self), 0(L-internal), 0(R-internal)
childL: 0x0, childR: 0x0
atomic: 0
---------------------------
\end{lstlisting}
For internal nodes,
\begin{itemize}
\item There are nodes such as 16382 where all attributes are defined.
\item However, there are nodes such as 16383 which are ``half-filled". Information has propagated up from its left child, however it is not yet connected to its right child. Hence, the atomic counter has value 1.
\item In fact, there are nodes such as 16384 which are yet to be populated in any way. For a binary radix tree, \plaincode{n\_triangles - 1} internal nodes should always be used.
\item Finally, the last internal node (with index \plaincode{29259 = n\_triangles - 1}) should contain the address of the root node in \plaincode{childL}. This can only happen if bottom-up construction reaches the top.
\item Having a NULL pointer for the root node would prevent the BVH from being traversed.
\end{itemize}

\begin{lstlisting}[backgroundcolor=\color{white}, basicstyle={\ttfamily\footnotesize}, escapechar=@, frame=single, linewidth=165mm]
Leaf nodes
   :
[16383] x:[992.584,1018.47], y:[363.936,387.745], z:[61.2898,62.1155]
self: 0xe2335ffa0, parent: 0xe2361ff40
triangleID: 4345

[16384] x:[968.359,992.584], y:[341.158,363.936], z:[61.0106,63.3395]
self: 0x0, parent: 0x0
triangleID: 4077
   :
\end{lstlisting}
For leaf nodes,
\begin{itemize}
\item There are nodes such as 16384 that are not connected with its parent.
\item This means \plaincode{\_\_device\_\_ void bvhUpdateParent(BVHNode* node, BVHNode* internalNodes, BVHNode* leafNodes, MORTON* morton, int nNodes)} has not been performed on this leaf node, even though it has been initialised with a triangleID by \plaincode{kernelBVHReset}.
\end{itemize}

\subsubsection{The culprit}\label{sec:culprit2}
As \sourcecode{bvhUpdateParent} is only recursively called by itself or \sourcecode{kernelBVHConstruct}, the evidence points toward a bug with the invocation of \sourcecode{self.kernel\_bvh\_construct}. Inspection of \plaincode{pycuda\_ray\_surface\_intersect.py} reveals the {\color{ruby}grid argument} (number of thread-blocks) {\color{ruby}was incorrectly configured} for this kernel.
\begin{lstlisting}[backgroundcolor=\color{gray1}, basicstyle={\ttfamily\footnotesize}, escapechar=@, frame=single, linewidth=165mm]
self.kernel_bvh_construct(
            self.d_internalNodes, self.d_leafNodes, self.d_morton,
            np.int32(self.n_triangles), block=self.block_dims, grid=self.@{\color{ruby}grid\_lambda}@)
\end{lstlisting}

It should read
\begin{lstlisting}[backgroundcolor=\color{gray1}, basicstyle={\ttfamily\footnotesize}, escapechar=@, frame=single, linewidth=165mm]
self.kernel_bvh_construct(
            self.d_internalNodes, self.d_leafNodes, self.d_morton,
            np.int32(self.n_triangles), block=self.block_dims, grid=self.@{\color{blue}grid\_dimsT}@)
\end{lstlisting}
based on the definitions
\begin{lstlisting}[backgroundcolor=\color{gray1}, basicstyle={\ttfamily\footnotesize}, escapechar=@, frame=single, linewidth=165mm]
self.grid_lambda = (self.grid_xLambda, 1)
self.grid_dimsT = (int(np.ceil(self.n_triangles / self.block_x)), 1)
\end{lstlisting}

When the program is executed, the console also shows
\begin{lstlisting}[backgroundcolor=\color{white}, basicstyle={\ttfamily\footnotesize}, escapechar=@, frame=single, linewidth=165mm]
CUDA partitions: 1024 threads/block,
grids: [rays: 9766, bvh_construct: 29, bvh_intersect: 16]
\end{lstlisting}
That is, \plaincode{grid\_lambda = (16, 1)} and \plaincode{grid\_dimsT = (29, 1)}. As \plaincode{kernel\_bvh\_construct} relates to triangle-mesh tree construction, the partitions should depend on \plaincode{grid\_dimsT} (equivalently, \plaincode{n\_triangles}). The number of partitions should not depend on \plaincode{grid\_lambda} which corresponds to the $N_\text{grids}$ parameter which relates to \plaincode{\_\_device\_\_ void bvhFindCollisions} and \plaincode{\_\_global\_\_ void kernelBVHIntersection}. This was a copy-and-paste error.

\subsubsection{Resolution}\label{sec:resolution2}
For this test case, \plaincode{grid\_lambda} < \plaincode{grid\_dimsT}. This explains why some leaf nodes were untouched and the associated data (volume bounds and indices) did not propagate to the top of the tree. It was doing less work than it was supposed to. The final PyCUDA implementation contains the bug-fix indicated in blue above.

\vspace{3mm}
\section{Summary}
These examples demonstrate that PyCUDA can make kernel code easier to debug. It is possible to transfer data from device memory to host with ease and inspect objects using PyCUDA abstraction and standard Python tools. For more complex C structures such as the bounding volume hierarchy, a decoding layer/method is required to interpret mixed-type content in individual \plaincode{BHVNodes} that constitute the internal and leaf node arrays. With such power also comes responsibility. There are certain pitfalls that developers ought to be mindful of. In the first case study, one is reminded of the different interpretations of \plaincode{int}\!---\,the Nvidia CUDA compiler (nvcc) treats int as int32 whereas numpy uses int64. This discrepancy creates a mismatch situation (see Fig.~\ref{fig:int-pointer}) where the int32* pointer lags behind the actual int64 data. This resulted in the ``disappearance'' of mesh triangles from the binary radix tree and misdetection of certain ray-surface intersections in the test scenario. This problem could have manifested in other ways and led to other strange behaviour in a different context. In the second case study, we encountered an exception. The same debugging strategy was employed and we managed to find the root cause of the exception. We found that an incorrect specification of the grid size contributed to an incomplete construction of the BVH. Inspection of the BVH structure clearly revealed some leaf node properties (volumetric bounds) had not propagated to the top of the tree. The final PyCUDA ray-surface intersection library is free of these issues. The source code is available at \url{https://github.com/raymondleung8/gpu-ray-surface-intersection-in-cuda/tree/main/pycuda} under a BSD-3 clause license.

\newpage\section*{Epilogue}
Currently, the entire implementation uses 32-bit float. Inside the \!\plaincode{intersectMoller}\! function, rounding errors associated with the dot and cross-product calculations can accumulate. For edge cases where the ray runs parallel or almost parallel to the face of a triangle, the loss of precision can sometimes make a difference. In our experience, this is relatively uncommon---the incidence rate is about $10^{-6}$ [this depends on the number of intersection tests that are nearly degenerate]. If this is an issue, \plaincode{PyCudaRSI.\_\_init\_\_(params)} may be initialised with \plaincode{params[`USE\_DOUBLE\_PRECISION\_MOLLER']} set to \plaincode{True}\!. This will activate \plaincode{COMPILE\_DOUBLE\_PRECISION\_MOLLER} in \plaincode{pycuda\_source.py} and store intermediate quantities within \plaincode{\_\_device\_\_ int intersectMoller} using 64-bit floats.

\section*{Further reading}
GPU programming and CUDA concepts are covered by Hwu et al.\,in \cite{kirk2016programming}. This book is a valuable resource for beginners. Scripting-based approach to GPU run-time code generation (particularly PyCUDA) is discussed in \cite{kloeckner-gpu-rtcg09,kloeckner-gpu-scripting-pycuda13}. Bounding volume hierarchies and linear binary radix tree construction are described by Apetrei in \cite{apetrei2014fast}. Ray-triangle intersection tests are surveyed and evaluated by Jim{\'e}nez et al. in \cite{jimenez2010robust,jimenez2014performance}.

\section*{Acknowledgements}
This work has been supported by the Rio Tinto Centre for Mine Automation and Australian Centre for Field Robotics.

\bibliographystyle{unsrt}
\bibliography{pycuda_references.bib}

\begin{thebibliography}{10}

\bibitem{leung2022gpursi}
Raymond Leung.
\newblock {GPU} implementation of a ray-surface intersection algorithm in
  {CUDA}. {S}ource code available at
  \url{https://github.com/raymondleung8/gpu-ray-surface-intersection-in-cuda}
  under the {BSD} 3-clause license. {D}ocumentation available at:
  \url{https://arxiv.org/abs/2209.02878}.
\newblock {\em ar{X}iv e-print}, pages 1--11, 2022.

\bibitem{kloeckner-gpu-rtcg09}
Andreas Kl{\"{o}}ckner, Nicolas Pinto, Yunsup Lee, Bryan Catanzaro, Paul
  Ivanov, and Ahmed Fasih.
\newblock Pycuda: {GPU} run-time code generation for high-performance
  computing. {A}vailable at \url{http://arxiv.org/abs/0911.3456}.
\newblock {\em arXiv e-print 0911.3456}, 2009.

\bibitem{kloeckner-gpu-scripting-pycuda13}
Andreas Kl{\"{o}}ckner, Nicolas Pinto, Bryan Catanzaro, Yunsup Lee, Paul
  Ivanov, and Ahmed Fasih.
\newblock {GPU} scripting and code generation with {P}y{CUDA}. {A}vailable at
  \url{http://arxiv.org/abs/1304.5553}.
\newblock In Wen-mei Hwu, editor, {\em GPU Computing Gems, Chapter 27}.
  Elsevier, 2011.

\bibitem{moller1997fast}
Tomas M{\"o}ller and Ben Trumbore.
\newblock Fast, minimum storage ray-triangle intersection.
\newblock {\em Journal of Graphics Tools}, 2(1):21--28, 1997.

\bibitem{karras2012maximizing}
Tero Karras.
\newblock Maximizing parallelism in the construction of {BVH}s, octrees, and
  k-d trees.
\newblock In {\em Proceedings of the Fourth ACM SIGGRAPH/Eurographics
  conference on High-Performance Graphics}, pages 33--37, 2012.

\bibitem{kloeckner-pycuda-doc}
Andreas Kl{\"{o}}ckner.
\newblock Py{CUDA} documentation. {A}vailable at:
  https://documen.tician.de/pycuda/, 2022.

\bibitem{nishino2017cupy}
Ryosuke Okuta, Yuya Unno, Daisuke Nishino, Shohei Hido, and Crissman Loomis.
\newblock Cu{P}y: A {N}um{P}y-compatible library for {NVIDIA} {GPU}
  calculations.
\newblock {\em 31st Confernce on Neural Information Processing Systems},
  151(7), 2017.

\bibitem{kirk2016programming}
Wen-Mei Hwu, David Kirk, and Izzat El~Hajj.
\newblock {\em Programming Massively Parallel Processors: A Hands-on Approach}.
\newblock Morgan Kaufmann, 2022.

\bibitem{apetrei2014fast}
Ciprian Apetrei.
\newblock Fast and simple agglomerative {LBVH} construction.
\newblock In Rita Borgo and Wen Tang, editors, {\em {EG} {UK} Computer Graphics
  \& Visual Computing}. The Eurographics Association. {A}vailable at:
  \url{http://diglib.eg.org/bitstream/handle/10.2312/cgvc.20141206.041-044/041-044.pdf?sequence=1&isAllowed=y},
  2014.

\bibitem{jimenez2010robust}
Juan~J Jim{\'e}nez, Rafael~J Segura, and Francisco~R Feito.
\newblock A robust segment/triangle intersection algorithm for interference
  tests. {E}fficiency study.
\newblock {\em Computational Geometry}, 43(5):474--492, 2010.

\bibitem{jimenez2014performance}
Juan~J Jim{\'e}nez, Carlos~J Og{\'a}yar, Jos{\'e}~M Noguera, and F{\'e}lix
  Paulano.
\newblock Performance analysis for {GPU}-based ray-triangle algorithms.
\newblock In {\em 2014 International Conference on Computer Graphics Theory and
  Applications ({GRAPP})}, pages 1--8. IEEE, 2014.

\end{thebibliography}

\end{document}